\documentclass[preprint]{aastex}
\newcommand{\kms}{km~s$^{-1}$}
\usepackage{natbib}
\usepackage{graphicx}
\usepackage{amssymb}
\usepackage{txfonts}
\begin{document}

\title{A recent accretion burst in the low-mass protostar
  IRAS~15398--3359: \\ ALMA imaging of its related chemistry}

\author{Jes K. J{\o}rgensen\altaffilmark{1,2}, Ruud
  Visser\altaffilmark{3}, Nami Sakai\altaffilmark{4}, Edwin
  A. Bergin\altaffilmark{3}, Christian Brinch\altaffilmark{1,2},
  Daniel Harsono\altaffilmark{5,6}, Johan E. Lindberg\altaffilmark{2,1},
  Ewine F. van Dishoeck\altaffilmark{5,7}, Satoshi
  Yamamoto\altaffilmark{4}, Suzanne E. Bisschop\altaffilmark{2,1},
  Magnus V. Persson\altaffilmark{5}}

\altaffiltext{1}{Niels Bohr Institute, University of Copenhagen,
  Juliane Maries Vej 30, DK-2100 Copenhagen {\O}., Denmark, jeskj@nbi.dk}
\altaffiltext{2}{Centre for Star and Planet Formation \& Natural History Museum of Denmark, University of Copenhagen, {\O}ster Voldgade 5--7, DK-1350 Copenhagen
  {K}., Denmark}
\altaffiltext{3}{Department of Astronomy, University of Michigan, 500 Church Street, Ann Arbor, MI 48109-1042, USA}
\altaffiltext{4}{Department of Physics, The University of Tokyo, 7-3-1 Hongo, Bunkyo-ku, Tokyo, 113-0033, Japan}
\altaffiltext{5}{Leiden Observatory, Leiden University, PO Box 9513, NL-2300 RA Leiden, The Netherlands}
\altaffiltext{6}{SRON Netherlands Institute for Space Research, PO Box 800, 9700 AV, Groningen, The Netherlands}
\altaffiltext{7}{Max-Planck Institut f\"ur extraterrestrische Physik, Giessenbachstrasse, D-85748 Garching, Germany}

\begin{abstract}
  Low-mass protostars have been suggested to show highly variable
  accretion rates through-out their evolution. Such changes in
  accretion, and related heating of their ambient envelopes, may
  trigger significant chemical variations on different spatial scales
  and from source-to-source. We present images of emission from
  C$^{17}$O, H$^{13}$CO$^+$, CH$_3$OH, C$^{34}$S and C$_2$H toward the
  low-mass protostar IRAS~15398--3359 on 0.5$''$ (75~AU diameter)
  scales with the Atacama Large Millimeter/submillimeter Array (ALMA)
  at 340~GHz. The resolved images show that the emission from
  H$^{13}$CO$^+$ is only present in a ring-like structure with a
  radius of about 1--1.5$''$ (150--200~AU) whereas the CO and other
  high dipole moment molecules are centrally condensed toward the
  location of the central protostar. We propose that HCO$^+$ is
  destroyed by water vapor present on small scales. The origin of this
  water vapor is likely an accretion burst during the last
  100--1000~years increasing the luminosity of IRAS~15398--3359 by a
  factor of 100 above its current luminosity. Such a burst in
  luminosity can also explain the centrally condensed CH$_3$OH and
  extended warm carbon-chain chemistry observed in this source and
  furthermore be reflected in the relative faintness of its compact
  continuum emission compared to other protostars.
\end{abstract}

\keywords{astrochemistry --- ISM: abundances --- ISM: individual
  (IRAS~15398--3359) --- ISM: molecules --- stars: formation ---
  stars: protostars}

\maketitle
\clearpage
\section{Introduction}\label{introduction}
The discoveries of complex organic (and even prebiotic) molecules on
small scales of low-mass protostars
\citep{bottinelli04iras16293,iras2sma,jorgensen12} as well as the
presence of rotationally supported disks relatively early in their
evolution \citep{tobin12l1527,murillo13,lindberg13alma} sets an
interesting frame for the question: what are the initial conditions
for the chemistry in young solar system analogs? One of the key
challenges is to follow the chemistry with increasing temperature and
density as material falls in from molecular clouds to solar-system
scales close to the young star.

The chemistry leading to the formation of larger molecules likely
involves a complex balance between molecules freezing out on dust
grains, grain surface chemistry and eventual evaporation of the
species before they may be (re-)incorporated in ices in the
circumstellar disks \citep[see, e.g.,][for recent
reviews]{herbst09,caselli12}. However, not only does this depend on
the wide-ranging network of chemical reactions -- but more
fundamentally also on the exact physical history of material as it is
accreted. In the simplest picture the temperature increases
monotonically as material falls in closer to the central protostar at
a constant rate. In reality, however, variations in, for example, the
accretion rate may strongly affect the protostellar luminosity and
thus temperature throughout the cloud. Since the dust temperature is
crucial for the gas-grain chemistry, luminosity variations can have a
strong effect on the radial distribution of the chemistry -- and from
source to source.

With the high angular resolution and surface brightness sensitivity of
the Atacama Large Millimeter/submillimeter Array (ALMA), it is
becoming possible to image the distribution of molecular species
toward individual protostars on Solar System scales. This paper
presents ALMA images of the deeply embedded protostar IRAS~15398--3359
(IRAS15398 hereafter), located in the Lupus I molecular cloud (B228)
\citep{heyer89,chapman07} at a distance of 155~pc
\citep{lombardi08}. From a reanalysis of its SED and submillimeter
continuum maps, its luminosity $L_{\rm bol}$=1.8~$L_\odot$, bolometric
temperature, $T_{\rm bol}$=44~K and envelope mass, $M_{\rm env}\approx
1.2~M_\odot$ makes it a fairly standard ``Class 0'' low-mass
protostar. From a chemical point of view IRAS15398 is interesting as
it is one of two prototypical sources with prominent carbon-chain
molecules such as C$_4$H, C$_4$H$_2$, CH$_3$CCH and HC$_5$N
\citep{sakai09}. This peculiar chemistry has been suggested to be
either an indication of the primordial conditions of the clouds and
relative time-scales and degrees of freeze-out for CO and CH$_4$
\citep{sakai09} or a flatter density and temperature profile on larger
scales of the envelope \citep[e.g.,][]{herbst09}. The main aim of this
paper is to relate the observed chemical emission signatures to the
physical structure and evolution of the protostar.

\section{Observations}
IRAS~15398--3359 was observed on four occasions between 2012 March 29
and 2012 April 11 as part of the ALMA Early Science Cycle~0 program
2011.0.00628.S (PI Jes J{\o}rgensen) at 0.87~mm (ALMA band 7). At the
time of observations 15--16 antennae were present in the array in an
extended configuration providing baselines ranging from 20--392~m
($\approx$~23--445~k$\lambda$). The phase center was taken to be
$\alpha$=15$^{\rm h}$43$^{\rm m}$02\fs16; $\delta$=$-$34\degr09\arcmin06\farcs80
[J2000]. Over the four observing sessions IRAS15398 was observed with
a total on-source integration time of about three hours.

The observations contain four spectral windows with 3840 channels and
a channel width of 122~kHz (0.11~\kms). The spectral setup was chosen
to cover 336.95--337.45~GHz, 338.30--338.80~GHz, 349.35--349.85~GHz
and 346.90--347.40~GHz. A second dataset obtained in connection with
this program, targeting the protostar R~CrA-IRS7B with a similar
spectral setup, is presented by \cite{lindberg13alma}.

The reduction followed standard recipes in CASA\footnote{\tt
  http://casa.nrao.edu/} \citep{mcmullin07} with calibration of the
complex gains through observations of the quasar J1517-243, passband
calibration on J1325-430 and flux calibration on Neptune and
Titan. This procedure provides a dataset with a beam size of about
0.55\arcsec$\times$0.37\arcsec\ (PA~=~$-$82$^\circ$), slightly varying
with wavelength, and an RMS level of
13~mJy~beam$^{-1}$~channel$^{-1}$. The continuum was constructed by
averaging the channels across the four spectral bands that appear
reasonably free of line emission with a resulting RMS level of about
0.28~mJy~beam$^{-1}$.

The CH$_3$OH $7_k-6_k$ transitions at 338.3--338.7~GHz were also
targeted in a parallel program (2011.0.00777.S; PI Nami Sakai) and
observed with 24 antennas for 1~hour on-source. For those CH$_3$OH
transitions as well as the continuum we combine the datasets for a
slight improvement in signal-to-noise.

\section{Results}
\begin{figure}
\resizebox{\hsize}{!}{\includegraphics{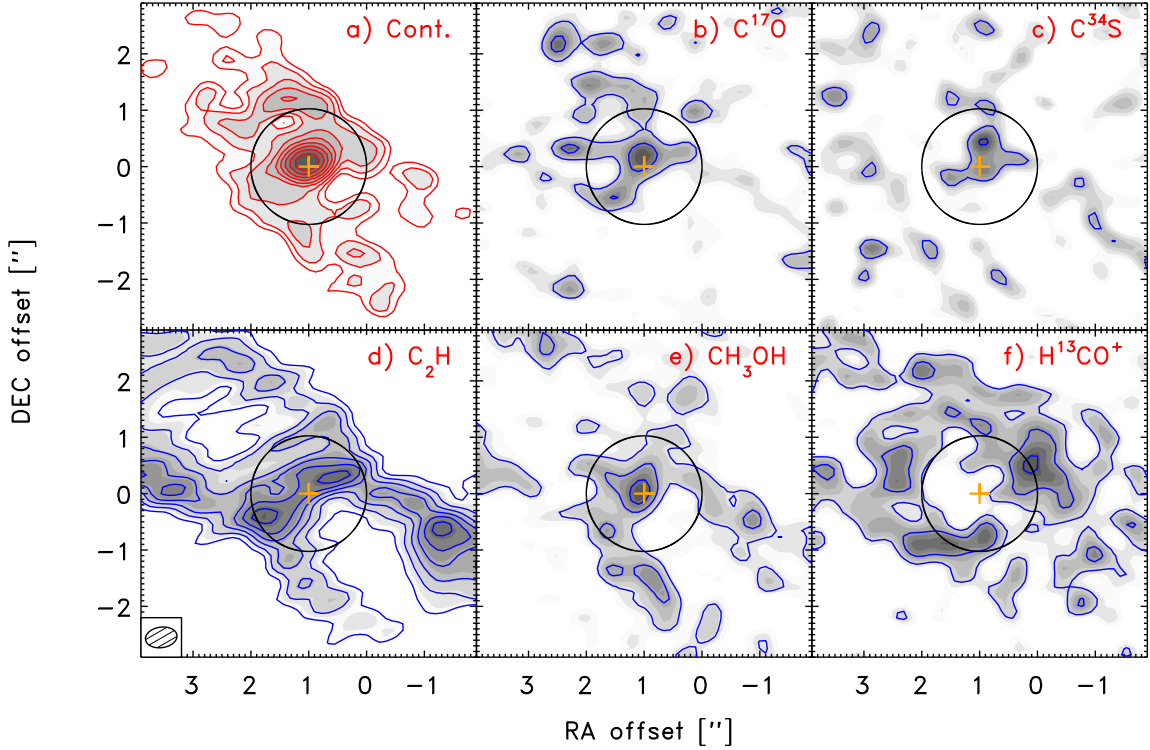}}
\caption{Maps of the continuum (\emph{a}) and integrated line emission
  toward IRAS15398 for (\emph{b}) C$^{17}$O $J=3-2$, (\emph{c})
  C$^{34}$S $J=7-6$, (\emph{d}) C$_2$H $N=4-3$, $J=7/2-5/2$,
  (\emph{e}) CH$_3$OH $J_k=7_0-6_0$ and (\emph{f}) H$^{13}$CO$^+$
  $J=4-3$. For the continuum the contours are given in 10 logarithmic
  steps from 4$\sigma$ (0.8~mJy~beam$^{-1}$) to the maximum
  (19~mJy~beam$^{-1}$). For each line, the emission is integrated over
  velocity intervals of $\pm $1~\kms\ from systemic velocity -- except
  for C$_2$H, which is integrated over $\pm $1.5~\kms\ around
  349.40~GHz to cover the two, $F = 4-3$ and $F = 3-2$, transitions of
  that species. The emission contours are shown in steps of 3$\sigma$
  (1$\sigma
  \approx$~5~mJy~beam$^{-1}$~km~s$^{-1}$).}\label{lineimage}\label{contimage}
\end{figure}

The continuum emission from IRAS15398 is clearly detected at
$\alpha$=15$^{\rm h}$43$^{\rm m}$02\fs24; $\delta$=$-$34\degr09\arcmin06\farcs71
[J2000] with a peak flux of 19~mJy~beam$^{-1}$ and integrated flux of
28~mJy (Fig.~\ref{contimage}a). For comparison the
JCMT/SCUBA image of IRAS15398 from the SCUBA legacy archive
\citep{difrancesco08} shows the core on larger scales with a total
integrated flux of 3.9~Jy at 850~$\mu$m. The bright part of the
continuum emission in the ALMA maps is largely unresolved. Some low
level extended emission is seen though, reflecting the resolved-out
low surface brightness emission from the ambient envelope seen in the
JCMT data. Generally, emission that is smooth on scales much larger
than 9$''$ (1400~AU) or present on scales larger than 16$''$ (2500~AU)
will be unreliable due to the lack of short baselines and size of the
ALMA primary beam at 345~GHz.

\begin{figure}
\resizebox{\hsize}{!}{\includegraphics{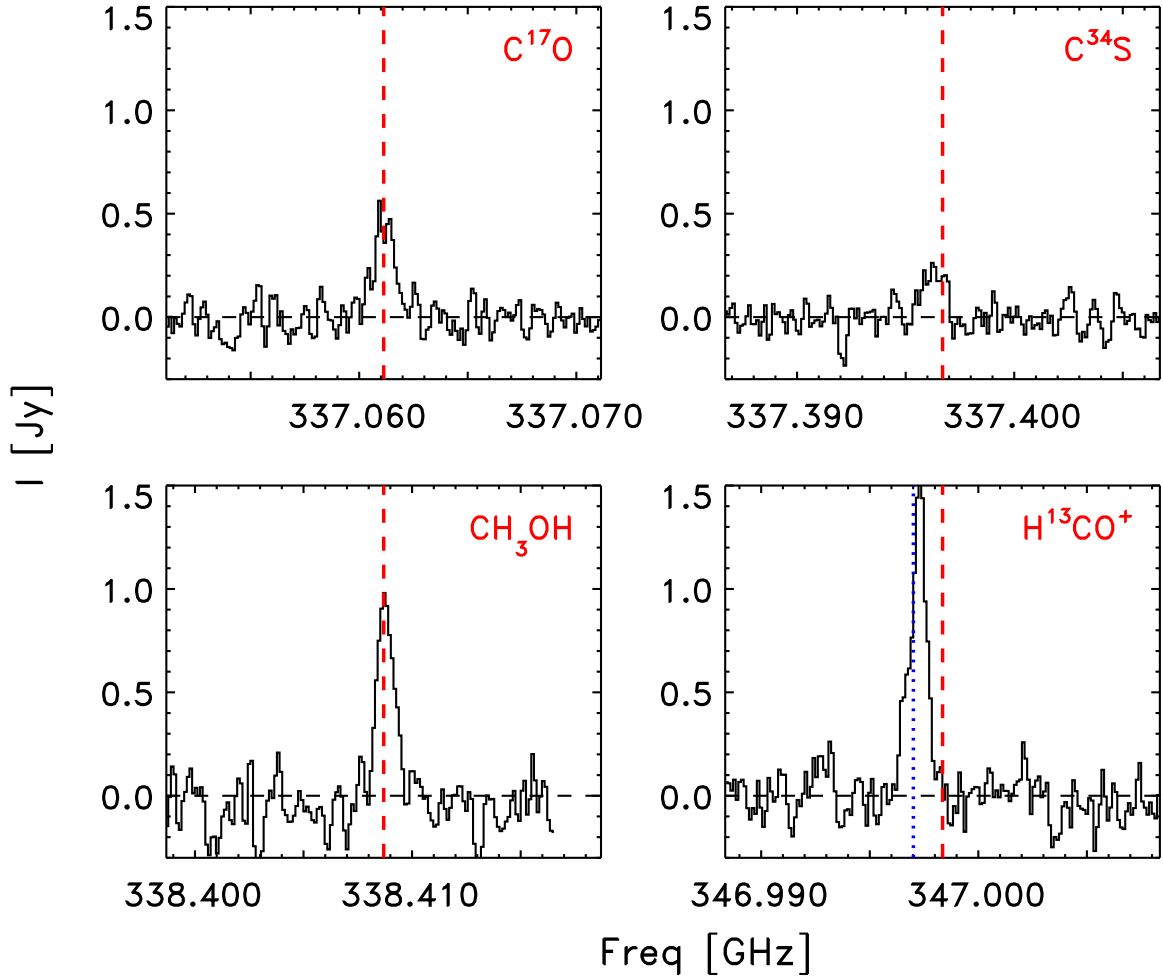}}
\caption{Spectra of C$^{17}$O 3--2, C$^{34}$S 7--6, CH$_3$OH
  $7_0-6_0$ and H$^{13}$CO$^+$ 4--3 obtained by integration over the
  inner 2$''$ (radius) toward the continuum peak of IRAS15398. Each
  spectrum has been corrected by a systemic velocity of 5.0~\kms. In
  each panel the vertical dashed (red) lines indicates the rest
  frequency of the given transition. In the H$^{13}$CO$^+$ panel, the
  dotted line indicates the frequency corresponding to a systemic
  velocity of 6.0~\kms.}\label{spectra}
\end{figure}

Figs.~\ref{lineimage} and \ref{spectra} show images and spectra for
the detected lines while Appendix~\ref{spectrum_appendix} includes a
full overview of the four spectral settings. Toward the continuum
position, transitions of five different species are detected:
C$^{17}$O $J=3-2$, C$^{34}$S $J=7-6$, C$_2$H $N=4-3$, $J=7/2-5/2$,
H$^{13}$CO$^+$ $J=4-3$ and a range of the CH$_3$OH $J_k=7_k-6_k$ transitions
at 338~GHz. In addition, there are a few lines that tentatively can be
assigned to CH$_3$CN and CH$_3$OH at 349.4--349.7~GHz. The most
noteworthy non-detection is SiO 8--7 suggesting that there is no
strong shock activity on small scales. The line-profiles are
relatively narrow (FWHM $\approx 0.5-0.7$~kms$^{-1}$) and
symmetric. All lines are detected at a systemic velocity of 5.0~\kms\
with the exception of H$^{13}$CO$^+$ 4--3 that peaks red-shifted by
about 1.0~\kms\ compared to the other species. The C$^{17}$O,
C$^{34}$S and CH$_3$OH emission is found toward the central continuum
peak and concentrated within about one arcsecond radius
(Fig.~\ref{lineimage}). H$^{13}$CO$^+$ in contrast is predominantly
present beyond this radius in a ring-like structure with a width of
about 1\arcsec. A more detailed discussion of the extended C$_2$H
emission related to the outflow is deferred to a separate publication
(Y.~Oya et al., in prep.). The detection of CH$_3$OH and tentative
detections of CH$_3$CN and CH$_3$OCH$_3$ mark the first discoveries of
these complex organic species toward one of the sources characterized
by the presence of the carbon-chain molecules.

The velocity offset and peculiar morphology of H$^{13}$CO$^+$ is
significant. The H$^{13}$CO$^+$ line profile itself is very symmetric
and no emission or absorption is seen at the lower LSR velocity of the
other species. Together with the extended continuum and C$_2$H
emission being present on-source and the lack of clear negative bowls
of emission, this suggests that spatial-filtering or limited dynamical
range due to the $(u,v)$-coverage is not the reason. Instead it
suggests that H$^{13}$CO$^+$ is present in gas that is kinematically
and morphologically different than the other species. With the current
sensitivity and resolution it is not possible to address the kinematic
nature of this component further, however.

\section{Discussion}
The most striking aspect of the data is the presence of C$^{17}$O and
lack of H$^{13}$CO$^+$ toward the central position in IRAS15398. An
explanation could be that the innermost region represents a lower
density environment where C$^{17}$O 3--2 could be excited and not the
higher density tracer H$^{13}$CO$^+$ 4--3. However, with the
additional detections of C$^{34}$S and CH$_3$OH on small scales this
can be ruled out. The differences between these species must be the
result of their chemistry.

At the densities and temperatures of the bulk protostellar envelope
material, the abundances of CO and HCO$^+$ are related through the
main formation and destruction mechanisms for HCO$^+$:
\begin{eqnarray*}
{\rm CO} + {\rm H}_3^+ & \rightarrow & {\rm HCO}^+ + {\rm H}_2 \\
{\rm HCO}^+ + {\rm e}^- & \rightarrow & {\rm CO} + {\rm H}
\end{eqnarray*}
Indeed, over large ranges of CO abundances seen in protostars, these
two species are observed to be strongly correlated, likely to first
order reflecting the chemistry driven by the freeze-out of CO
\citep{paperii}. The anti-correlation between CO and HCO$^+$ in our
images suggests that another process is at work as well. 

The observed profiles of C$^{17}$O and H$^{13}$CO$^+$ can be used to
quantify their abundances through line radiative transfer
modeling. For this purpose, we adopted a 1D model for the IRAS15398
envelope with a power-law density profile, constrained by the spectral
energy distribution and submillimeter continuum images, and with the
temperature distribution calculated self-consistently using the
Transphere code\footnote{\tt
  http://www.ita.uni-heidelberg.de/$\sim$dullemond/software/transphere/index.shtml}
\citep{dullemond02}. The IRAS15398 envelope is well-fit with a
power-law density profile decreasing toward larger radii as $n({\rm
  H}_2) \propto r^{-p}$ with $p = 1.5-2$ and a density of $(1-3)\times
10^7$~cm$^{-3}$ and a temperature of 30~K at 175--200~AU (1--1.5$''$).

\begin{figure}
\raisebox{2.5cm}{\resizebox{0.45\hsize}{!}{\includegraphics{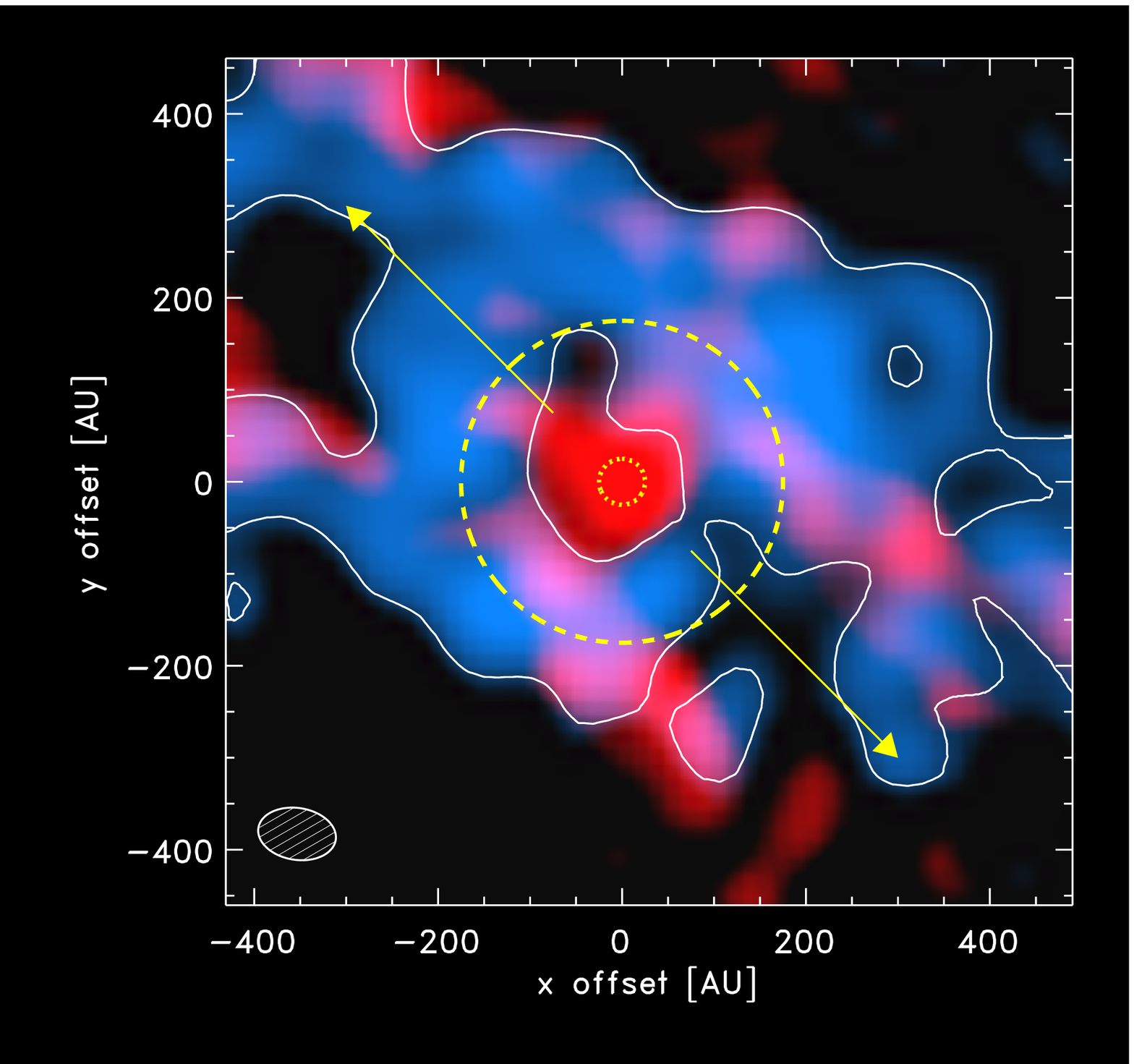}}}
\resizebox{0.5\hsize}{!}{\includegraphics{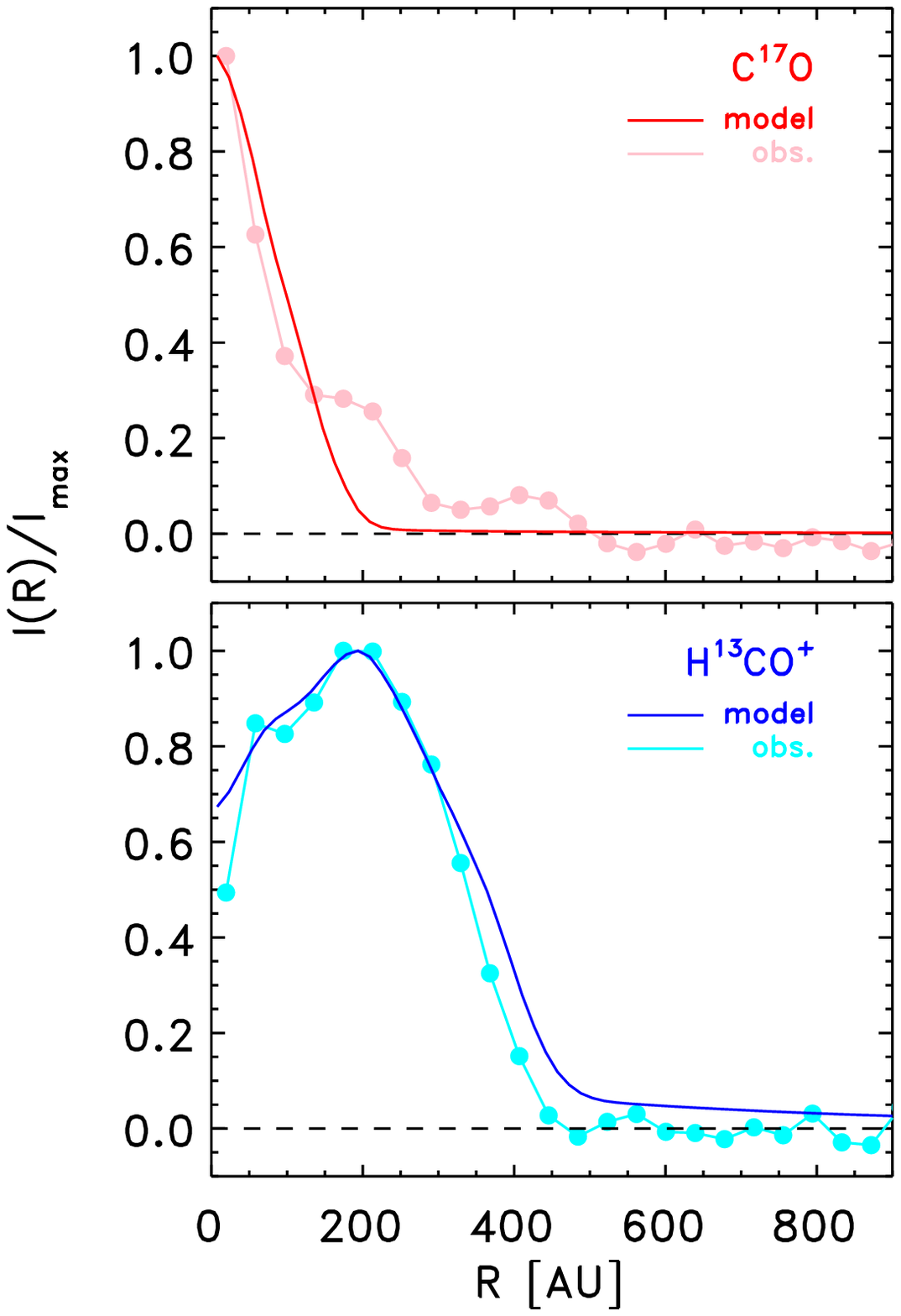}}
\caption{\emph{Left:} Observed emission of CH$_3$OH in red and
  H$^{13}$CO$^+$ (blue) compared to the physical scales of the source
  and with the observed beam shown in the bottom left corner. The
  dotted and dashed circles indicate the radii where the temperatures
  have dropped to 100 K and 30 K, respectively according to the dust
  radiative transfer model for the envelope. The arrows indicate the
  propagation direction of the IRAS15398 outflow. \emph{Right:}
  Observed radial profiles of C$^{17}$O 3--2 and H$^{13}$CO$^+$ 4--3
  compared to radiative transfer models of both species. For C$^{17}$O
  the abundance changes from 5$\times 10^{-8}$ (a CO abundance of
  1$\times 10^{-4}$) in the inner regions of the envelope to 5$\times
  10^{-10}$ at the radius where the temperature drops below 30~K
  (i.e., depletion by a factor 100). For H$^{13}$CO$^+$ the abundance
  is 7$\times 10^{-11}$ in the region of the envelope where the
  temperature is between 20~K and 30~K. Inside and outside this region
  the H$^{13}$CO$^+$ abundance must be at least a factor 20
  lower.}\label{chemfits}
\end{figure}

Subsequently the line radiative transfer was performed for C$^{17}$O
and H$^{13}$CO$^+$ using the one-dimensional RATRAN code
\citep{hogerheijde00vandertak} following the same approach as in,
e.g., \cite{l483art}: the abundances of the two species were taken to
be constant in regions separated at specific temperatures and adjusted
to match the observed brightness profiles (Fig.~\ref{chemfits}). It is
possible to reproduce the observed emission for C$^{17}$O with an
abundance of 5$\times 10^{-8}$ (a CO abundance of 1$\times 10^{-4}$)
in the inner regions of the envelope with temperatures above 30~K
(175~AU) and 5$\times 10^{-10}$ at larger radii (i.e., depletion by a
factor 100). This confirms the expectation that C$^{17}$O is present
in the inner regions of the envelopes where the temperature increases
above the CO sublimation temperature -- similar to what is seen in
other protostars \citep[e.g.,][]{l483art}. For H$^{13}$CO$^+$ in
contrast it is only possible to reproduce the observed profiles with
an abundance of 7$\times 10^{-11}$ (HCO$^+$ abundance of 5$\times
10^{-9}$) in the region of the envelope where the temperature is
between 20~K and 30~K and a drop in abundance of at least a factor 20
inside and outside this region.

\begin{figure}
\resizebox{\hsize}{!}{\includegraphics{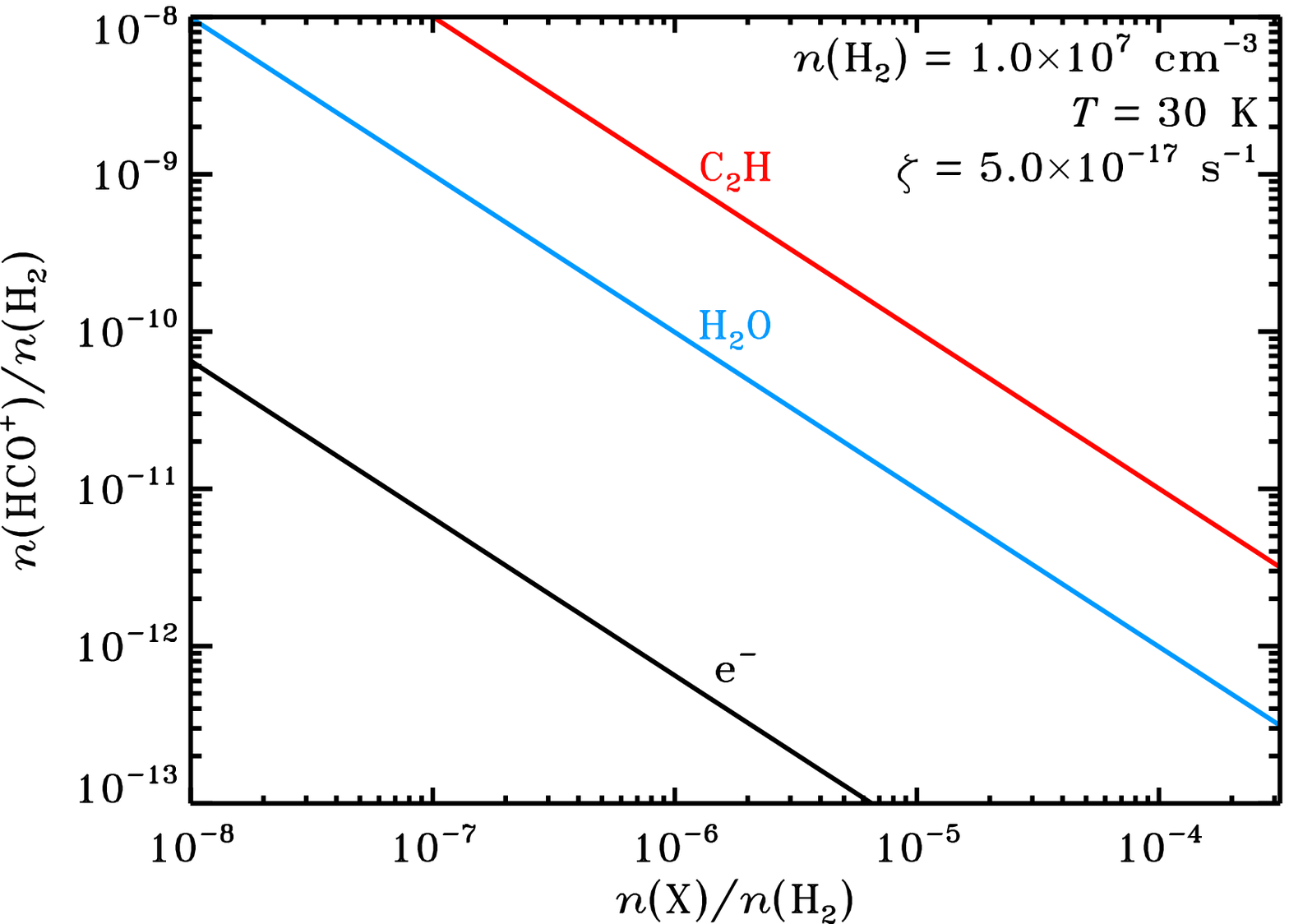}}
\caption{Abundance of HCO$^+$ as function of H$_2$O (blue) and C$_2$H
  (red) abundance in a gas with a density of $10^7$~cm$^{-3}$ and a
  temperature of 30~K in a chemical model based on
  \cite{visser12}. For water abundances as low as $10^{-6}$, the
  HCO$^+$ abundance is reduced by two orders of magnitude to the
  observed $\sim$10$^{-10}$. Similar levels of destruction by C$_2$H
  require an abundance of that species an order of magnitude larger,
  about 10$^{-5}$.}\label{watermodel}
\end{figure}

A possible explanation for the depletion of HCO$^+$ on small scales is
that it is destroyed by dipolar, neutral molecules. The most abundant
of these is H$_2$O, which predominantly is present in solid form in
protostellar envelopes. However, in regions where the temperatures
increase above 100~K water sublimates and becomes an important
destroyer of HCO$^+$. Fig.~\ref{watermodel} shows the result of a
steady state calculation based on the more detailed models of
\cite{visser12} utilizing the UDfA12 gas-phase chemical network
\citep{mcelroy13}. In this simple calculation we show the HCO$^+$
abundance as function of H$_2$O and C$_2$H abundance at a density of
$10^7$~cm$^{-3}$, temperature of 30~K and cosmic ray ionization rate
of $5\times 10^{17}$~s$^{-1}$. An H$_2$O abundance increase to
$10^{-6}$ is enough to reduce HCO$^+$ by two orders of magnitude --
matching the inferred abundance profile inferred from the
observations. Alternatively, other neutral species, such as the
carbon-chain molecules or H$_2$CO, could also work as destroyers of
HCO$^+$. However, due to their weaker dipole moments, their abundances
need to be comparably larger to destroy HCO$^+$ efficiently
(Fig.~\ref{watermodel}). For example, in the chemical models by
\cite{aikawa12}, H$_2$CO is the main destroyer of HCO$^+$ at
intermediate temperatures -- but the gas-phase abundances of H$_2$CO
in those models are as high as 10$^{-5}$, which are not supported by
observations of low-mass protostars that show typical abundances of
H$_2$CO $\lesssim$10$^{-7}$ in the warm gas
\citep[e.g.,][]{ceccarelli00b,schoeier02,hotcorepaper}.

The main issue about the scenario in which water acts as the main
destroyer is that temperatures of 100~K are only reached in the inner
25~AU for a 1.8~$L_\odot$ source such as IRAS15398 -- much smaller
than the H$^{13}$CO$^+$ ring. High water abundances on large scales
could be caused by, e.g., an outflow-driven shock -- and indeed lead
to the destruction of HCO$^+$ \citep[e.g.,][]{bergin98,i2art}. Water
is detected toward IRAS15398 with Herschel-PACS and found to be
associated with the outflow \citep{karska13}, but the ring-like
structure of H$^{13}$CO$^+$ and relatively narrow lines of CH$_3$OH
(another molecule that could be produced by ice sputtering) speak
against this being the reason for the observed emission structures
here. Another solution is that IRAS15398 has recently undergone a
change in luminosity -- e.g., related to a burst in accretion. Such
accretion bursts have been proposed to be characteristic of the
embedded protostellar stages and invoked to explain for example the
distribution of observed luminosities of embedded protostars
\citep[e.g.,][]{kenyon90,vorobyov06,evans09,dunham12}. An increase in
luminosity would sublimate water in a much larger region than hinted
by its current luminosity and after the burst water would stay in the
gas-phase for a period of time corresponding to the H$_2$O freeze-out
time-scales. To sublimate water at a distance of 175~AU of IRAS15398,
its luminosity (and accretion rate) should have been higher by two
orders of magnitude than currently. At the density of the IRAS15398
envelope at this distance of $10^7$~cm$^{-3}$, the time-scale for
water to freeze-out is of order $100-1000$~years, which can be
considered an upper limit to the time since the burst.

Unfortunately no direct imaging of the water vapor in IRAS15398 exists
-- but CH$_3$OH may work for a good proxy for water as it desorbs at
similar temperatures. Fig.~\ref{chemfits} illustrates that the
CH$_3$OH is indeed present and extended on the scales where
H$^{13}$CO$^+$ is absent, lending further credibility to this
suggestion. A simple estimate using \emph{Radex}
\citep{vandertak07radex} shows that the C$^{17}$O 3--2 and CH$_3$OH
$7_0-6_0$ are similar in strength for a [CH$_3$OH]/[CO] abundance
ratio of $6\times 10^{-5}$ at the density of $10^{7}$~cm$^{-3}$ and
temperature of 30~K characteristic of the CH$_3$OH emitting
region. Assuming a standard CO abundance of $\sim 10^{-4}$ this
translates into a CH$_3$OH abundance with respect to H$_2$ of $\sim
10^{-8}$ in this inner region.

The suggestion of a recent episodic accretion burst also raises
another interesting perspective -- namely it may have triggered the
chemistry leading to the formation of warm carbon-chain molecules on
large scales in the IRAS15398 envelope. Observations of the two
prototypical sources, L1527 and IRAS15398, show that these
carbon-chain molecules are present on large scales of a few thousand
AU but with a sharp abundance increase inward of 1000 AU
\citep{sakai08,sakai09}. \citeauthor{sakai08} mentioned that the
widespread emission is related to carbon-chain molecules produced in
the early phase of prestellar evolution -- but the additional increase
at 1000~AU scales require CH$_4$ being present in the gas-phase at
temperatures of $\approx$30~K \citep[see also][]{hassel08}. From the
model above the envelope temperatures of IRAS15398 is currently only
about 15~K on the 1000~AU scales and only reach 30~K in the inner
175~AU. However, if it has undergone a burst in recent history this
could have caused CH$_4$ to evaporate and trigger the carbon-chain
chemistry at much larger radii: an increase in luminosity by a factor
50--100 would for example shift the radius where the temperature is
30~K out to 1500--2200~AU.

It is interesting to note that L1527 and IRAS15398 both contain
relatively weak compact continuum emission compared to many other
deeply embedded protostars \citep[e.g.,][]{evolpaper}. With the
assumption in that paper, the observed compact continuum emission on
baselines $\gtrsim 50$~k$\lambda$ corresponds to 0.029~$M_\odot$ in
L1527 and $< 0.01$~$M_\odot$ in IRAS15398 -- compared to the median
0.089~$M_\odot$ for the Class 0 protostars. The nature of this compact
continuum emission is still unclear -- but one possibility is that it
reflects the build-up of an unstable (pseudo-)disk
\citep{evolpaper}. In that scenario, the absence of strong compact
continuum emission toward these two sources would indicate that they
have recently shedded these massive disks.

In summary, the proposed scenario of a recent accretion burst explains
four characteristic signatures observed toward IRAS15398: \emph{(i)}
the clear absence of HCO$^+$ in the region with prominent CO emission,
\emph{(i)} the presence of extended CH$_3$OH emission relative to its
current luminosity, \emph{(iii)} the presence of warm carbon-chain
molecules on large scales and \emph{(iv)} the absence of very massive
dust continuum component on small scales. Although, different
mechanisms can possibly explain any one of these signatures, the
strength of the proposed scenario is that it accounts for all. If the
scenario is confirmed through similar observations of other
protostars, statistics of the detections of such signatures could shed
further light onto the frequency of similar bursts during the embedded
protostellar stages.

\acknowledgments 

We thank the referee, Andrew Walsh, for a prompt and constructive
report. This paper makes use of ALMA dataset
\dataset{ADS/JAO.ALMA\#2011.0.00628.S}. ALMA is a partnership of ESO
(representing its member states), NSF (USA) and NIN~S (Japan),
together with NRC (Canada) and NSC and ASIAA (Taiwan), in cooperation
with the Republic of Chile. The Joint ALMA Observatory is operated by
ESO, AUI/NRAO and NAOJ.  This research was supported by a Lundbeck
Foundation Junior Group Leader Fellowship to Jes J{\o}rgensen. Centre
for Star and Planet Formation is funded by the Danish National
Research Foundation. We also acknowledge support from National Science
Foundation grant 1008800 and EU A-ERC grant 291141 CHEMPLAN.

\appendix
\section{Overview of full spectral windows}\label{spectrum_appendix}
\begin{figure}[!htb]
  \resizebox{\hsize}{!}{\includegraphics{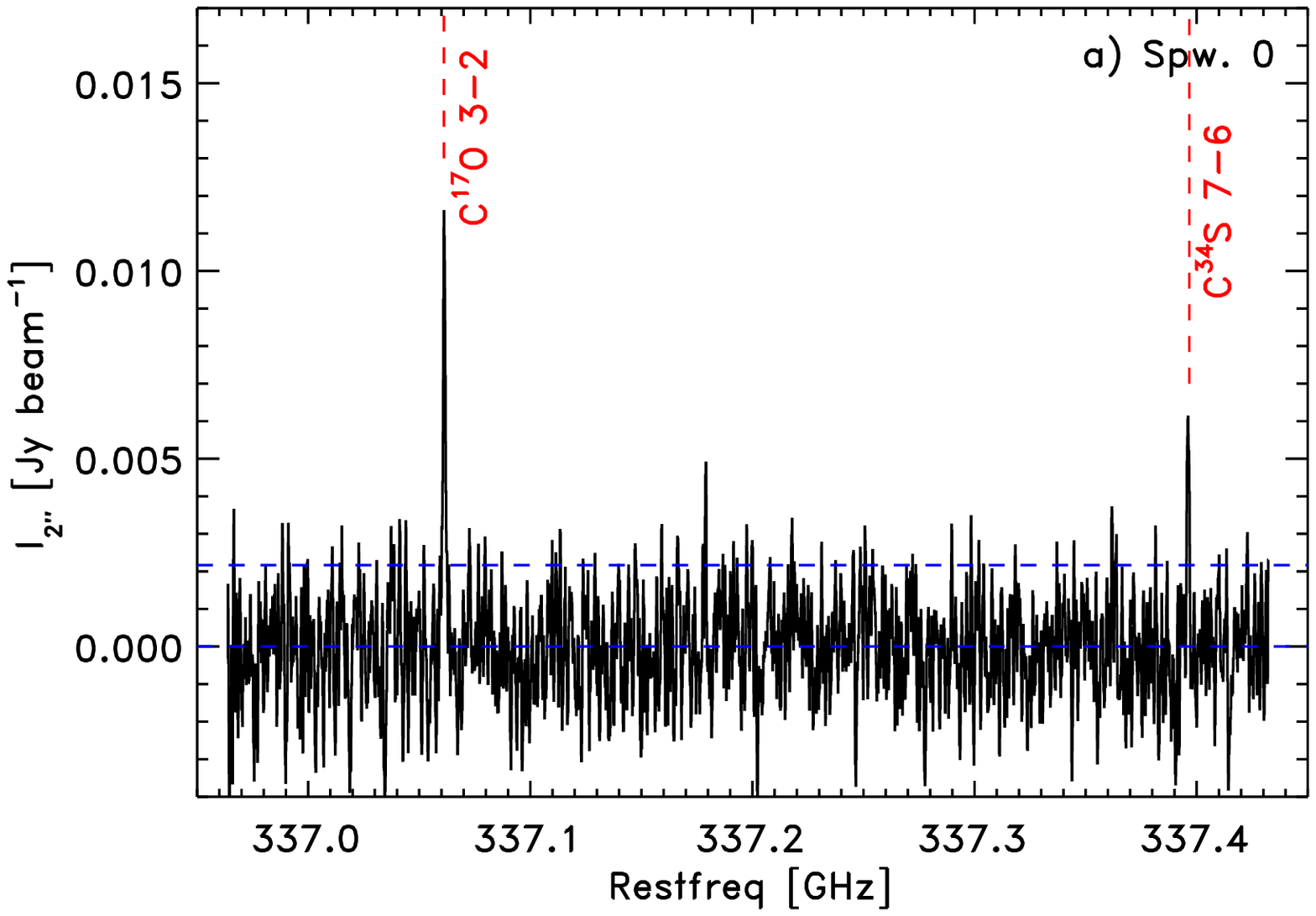}\includegraphics{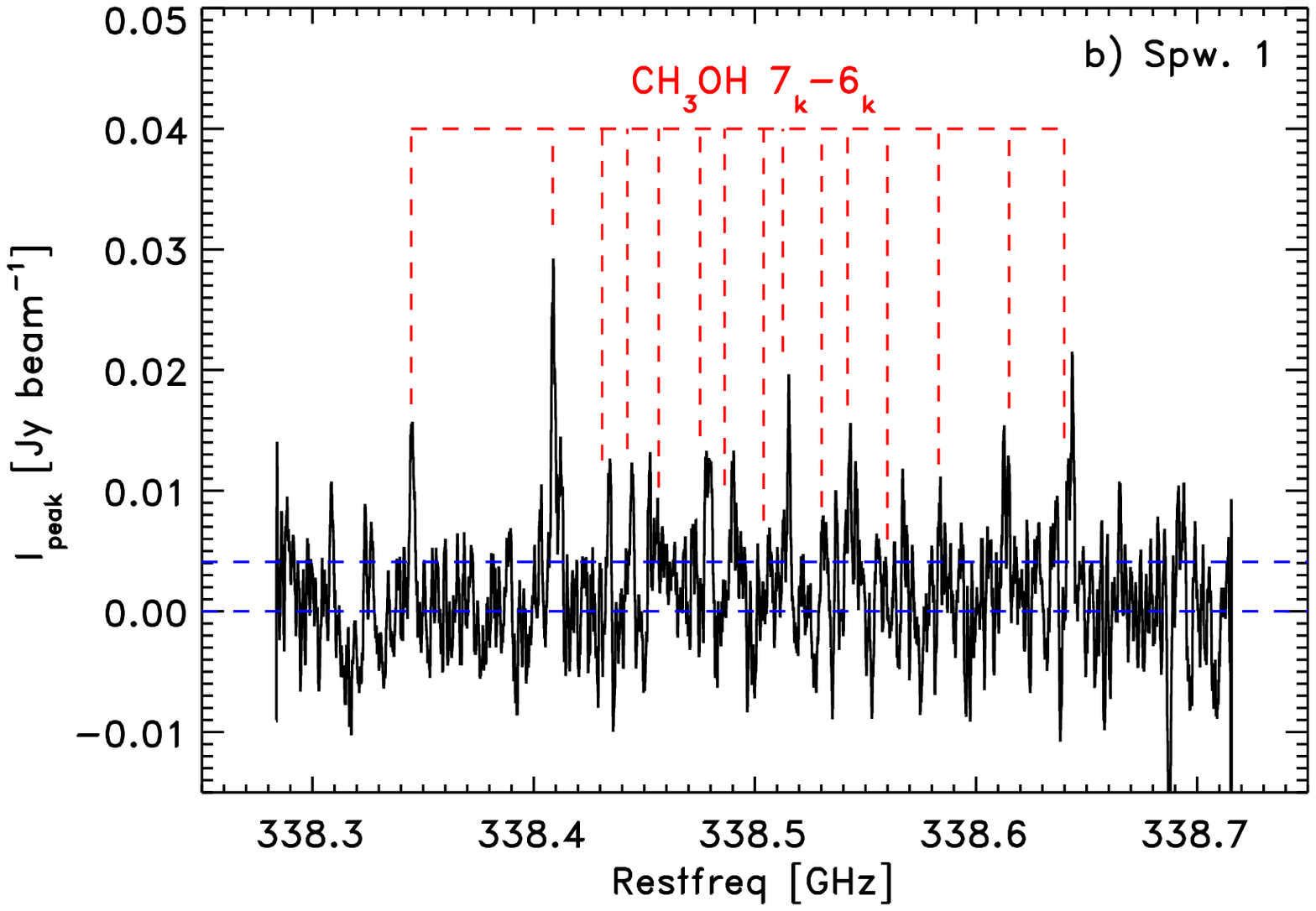}}
  \resizebox{\hsize}{!}{\includegraphics{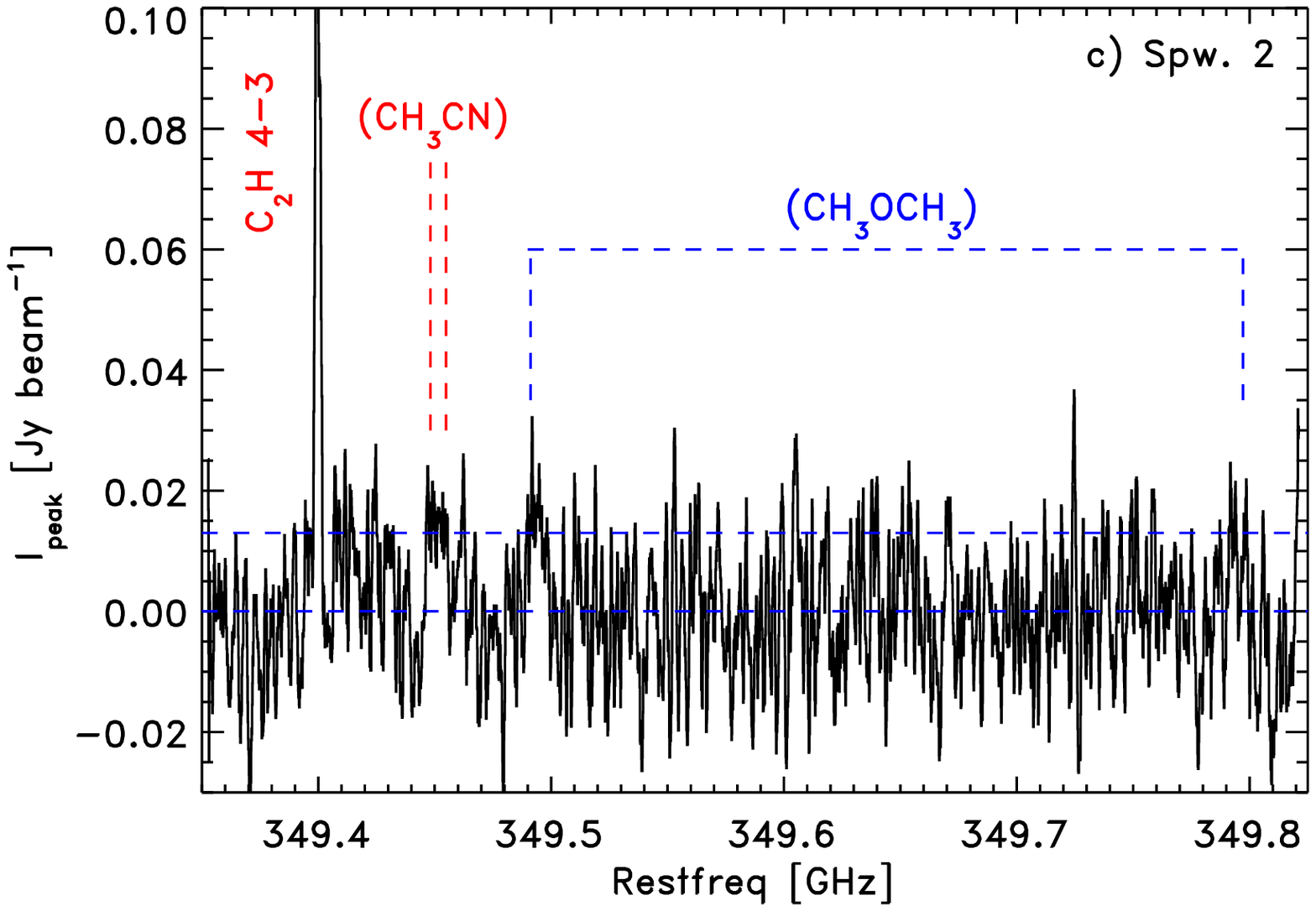}\includegraphics{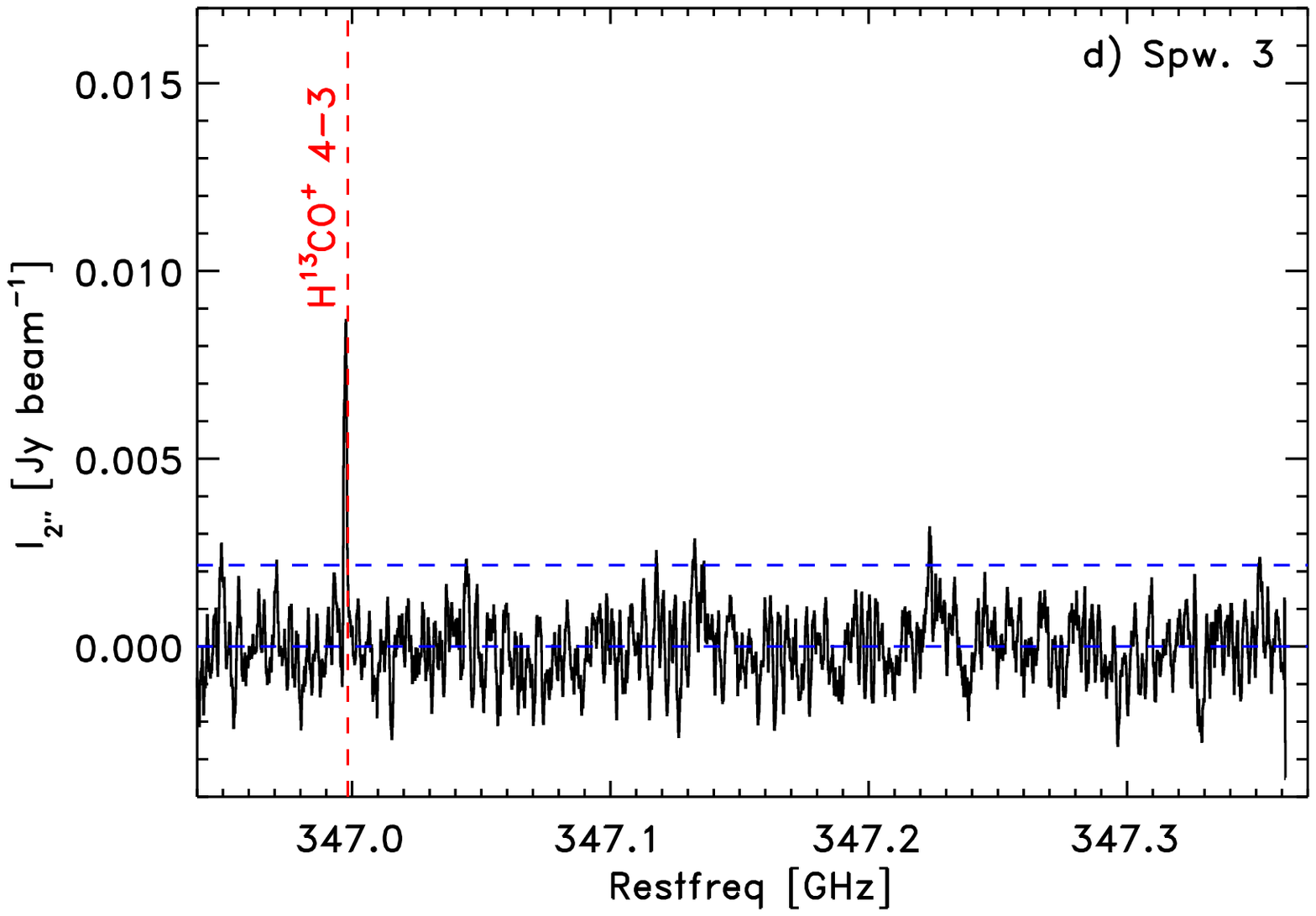}}
  \caption{Spectra from the four covered spectral windows taken in the
    central pixel toward the continuum peak (panels \emph{b} and
    \emph{c}) or averaged over the inner 2$''$ region (panels \emph{a}
    and \emph{d}). In panel \emph{b}) all cataloged CH$_3$OH $7_k-6_k$
    transitions have been marked -- no matter whether detected or
    not. }\label{spectrum}
\end{figure}

\pagebreak

\end{document}